\documentclass[american,aps,prl,reprint,groupedaddress,twocolumn,showpacs]{revtex4-1} 
\usepackage[unicode=true,pdfusetitle, bookmarks=true,bookmarksnumbered=false,bookmarksopen=false, breaklinks=false,pdfborder={0 0 0},backref=false,colorlinks=false] {hyperref}
\hypersetup{ colorlinks,linkcolor=myurlcolor,citecolor=myurlcolor,urlcolor=myurlcolor}
\usepackage{braket,colortbl,cleveref,amsthm,amsmath,amssymb,txfonts}
\definecolor{myurlcolor}{rgb}{0,0,0.7}
\usepackage{graphics,graphicx}
\usepackage{color}
\usepackage{colortbl}
\usepackage{subfigure}
\usepackage{graphicx}
\usepackage{ragged2e}
\usepackage{amsmath}
\usepackage{graphicx}
\usepackage[utf8x]{inputenc}
\usepackage{color}
\usepackage{amsmath}
\usepackage{tikz}
\usetikzlibrary{shapes.geometric, arrows}
\usepackage{latexsym}
\usepackage{amssymb}
\usepackage{amsthm}
\usepackage{bm}
\usepackage{graphics,epstopdf}
\usepackage{color}\usepackage{amsmath}

\usepackage[usenames,dvipsnames,svgnames]{pstricks}
\usepackage{epsfig}
\usepackage{pst-grad} 
\usepackage{pst-plot} 
\tikzstyle{startstop} = [rectangle, rounded corners, minimum width=3cm, minimum height=1cm,text centered, draw=black, fill=red!30]

\tikzstyle{env}=[circle,  ball color = green!20, minimum size= 80mm]
\tikzstyle{central}=[circle, ball color = red!100, minimum size=8mm]
\tikzstyle{bath}=[circle, ball color =blue!75, minimum size=4mm]

\theoremstyle{plain}

\def\bea{\begin{eqnarray}}
\def\eea{\end{eqnarray}}
\def\ba{\begin{array}}
\def\ea{\end{array}}

\def\beq{\begin{equation}}
\def\eeq{\end{equation}}

\begin{document}

\title{Superposition of causal order as a metrological resource for quantum thermometry}

\author{Chiranjib Mukhopadhyay}
\email{chiranjibmukhopadhyay@hri.res.in}
\author{Manish K. Gupta}
\email{manishh.gupta@gmail.com}
\author{Arun Kumar Pati}
\email{akpati@hri.res.in}
\affiliation{Quantum Information and Computation Group, Harish-Chandra Research Institute, Homi Bhabha National Institute, Allahabad 211019, India }


\begin{abstract}
\noindent  


We propose a novel approach to qubit thermometry using a quantum switch, that introduces an indefinite causal order in the probe-bath interaction, to significantly enhance the thermometric precision. The resulting qubit probe shows improved precision in both low and high temperature regimes when compared to optimal qubit probes studied previously. It even performs better than a Harmonic oscillator probe, in spite of having only two energy levels rather than an infinite number of energy levels as that in a harmonic oscillator. We thereby show unambiguously that quantum resources such as the quantum switch can significantly improve equilibrium thermometry. We also derive a new form of thermodynamic uncertainty relation that is tighter and depends on the energy gap of the probe. The present work may pave the way for using indefinite causal order as a metrological resource.


\end{abstract}

\maketitle

Low temperature sensing is of utmost importance in numerous instances ranging from many body physics \citep{why1,why3} to biophysics \citep{why2}. Quantum theory lends a very special status to parameters like temperature which cannot be represented by a Hermitian operator, therefore one has to estimate them through the measurement of other operators \citep{helstrom_book}. One of the challenges in thermometry lies in the fact that macroscopic probes may disturb the bath by distorting its thermal profile \citep{seveso}. Quantum thermometry \citep{sanpera_review,thermo_review, stace,qt1,qt2,wang,hofer, ivanov} is thus important, as it aims at improving the precision of nanoscale probes. In the case of metrology and parameter estimation, it is well known that spatial entanglement  \citep{nielsen_book,horodecki_rev}  between two distinct parties sometimes allows for a better scaling than the so called $standard$ $quantum$ $limit$ \citep{gio_review}. However, it has now been realized that quantum mechanics also allows for operations with superposition of causal order \citep{aharonov, chiribella_PRA, oreshkov, brukner, chiribella_review}. This idea has been recently exploited, among others, towards enhancing the classical capacity of channels \citep{ebler,manik}, reducing  communication complexity of tasks \citep{chiribella_PRA}, and improving teleportation protocols in noisy scenarios \footnote{C. Mukhopadhyay, A. K. Pati, \ \emph{in preparation}}. Experimental implementations of quantum switches have also been achieved \citep{Rubino, branciard} using optical setups. Thus, an inevitable question arises, can we get a metrological advantage in the presence of superposition of causal order?

\noindent In this letter, we provide an affirmative answer for the case of qubit thermometry. We show that, by using a quantum switch it is possible to estimate the temperature of a bath significantly more precisely than previously considered \emph{optimal qubit probes} \citep{correa_prl,sanpera_review} . While an optimal conventional qubit probe is outperformed by a Harmonic oscillator probe with infinite levels, we show that the same qubit probe, augmented with a quantum switch, can outperform the conventional Harmonic oscillator probe in the operating temperature window. We also derive thermodynamic uncertainty relations in the presence of the quantum switch. Our results open up the possibility of using indefinite causal order as a resource in  quantum metrology.

\emph{Preliminaries -} We first briefly review the existing theory of optimal qubit thermometry following \citep{correa_prl,sanpera_review}. The imprecision in estimating the inverse temperature $\beta$ from a probe which has attained equilibration in a thermal bath of inverse temperature $\beta$ is bounded from below by the quantum Cramer Rao bound  which assumes the form \begin{equation}
\delta \beta \geq \frac{1}{\sqrt{\nu \ \mathcal{F}_{\beta}}},
\label{tur}
\end{equation}
\noindent with $\nu$ being the number of measurements, and $\mathcal{F}_{\beta}$ being the $quantum$ Fisher information (QFI) of the thermalized probe state $\rho$, is given by \citep{paris_review}
\begin{equation}
\mathcal{F}_{\beta} (\rho) = \sum_{k} \frac{(\partial_{\beta} p_{k})^2}{p_k} + 2 \sum_{n\neq m} \frac{(p_{m} - p_{n})^2}{p_m+p_n} |\langle \psi_n|\partial_{\beta} \psi_m\rangle |^2 ,
\label{qfi_defn}
\end{equation}\noindent where $\lbrace p_k \rbrace$, and $\lbrace |\psi_k \rangle \rbrace$ are the eigenvalues and eigenvectors of the state $\rho$. For a single copy of the probe with a Hamiltonian $H$, the QFI equals the variance $\Delta H^2$ of the Hamiltonian. Thus, the above expression amounts to the following thermodynamic uncertainty relation 
\begin{equation}
\delta \beta \Delta H \geq 1
\end{equation}
\noindent For \emph{optimal} quantum thermometry in case of a qubit probe, one optimizes the QFI for the temperature over the parameter $x = \epsilon/T$, where $\epsilon$ is the energy gap of the probe Hamiltonian, and $T$ being the bath temperature, and thus obtains the following transcendental equation for $x = x^*$ \citep{correa_prl}

\begin{equation}
e^{x^*} = \frac{x^*+2}{x^*-2}  .
\label{opt_eqn}
\end{equation}

\noindent The above equation can be numerically shown to have the solution $x^{*} \approx 2.399 $. The resulting QFI for temperature has a peak, which determines the operating window of the thermometer. \\

\begin{figure}
\includegraphics[scale=0.25]{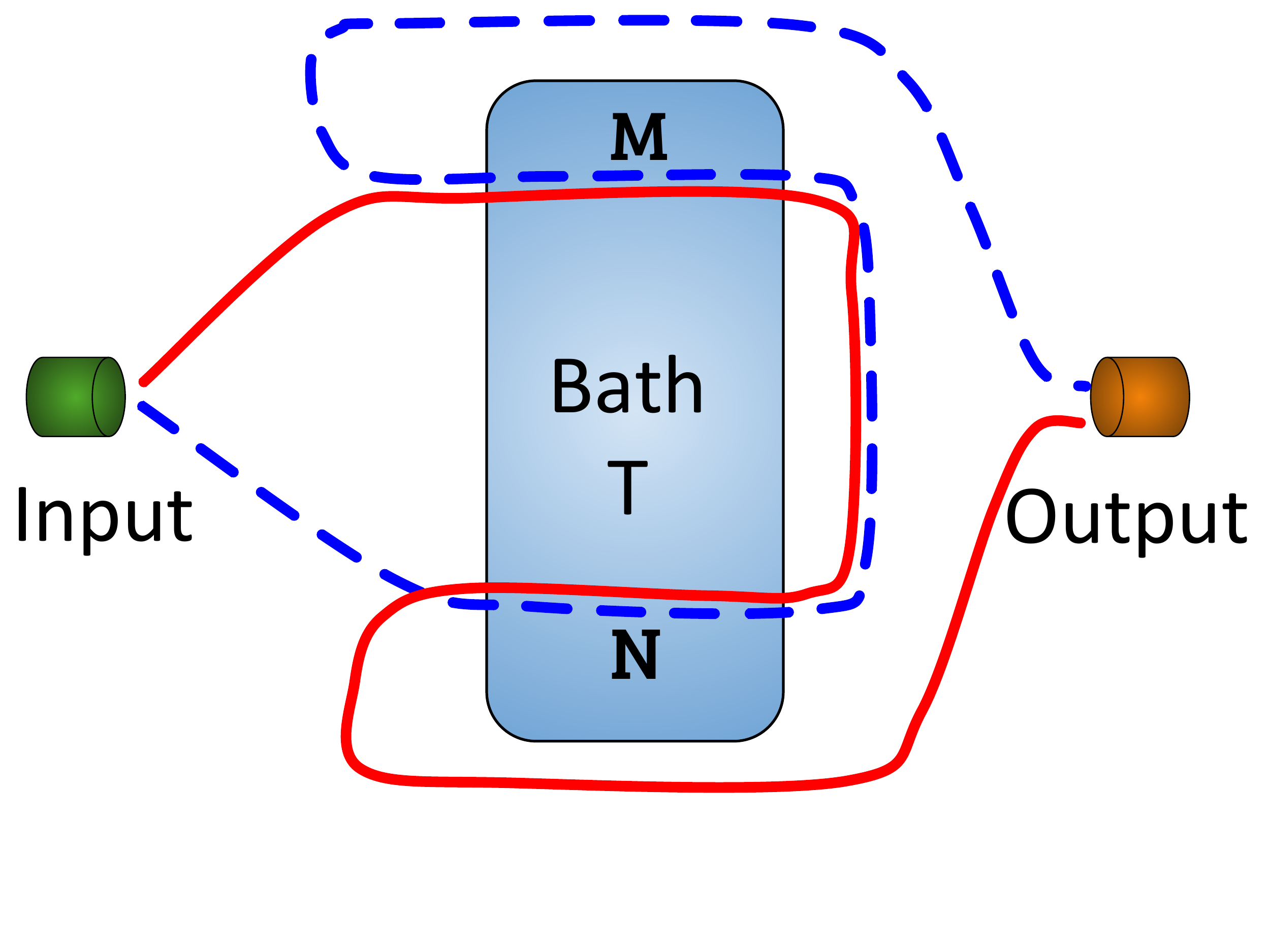}
\caption{(color online) Schematic of the protocol. The input probe is being sent in two possible pathways (red solid line, which implements $N \circ M$ and blue dashed line, which implements $M \circ N$) based upon the configuration of the control. $N$ and $M$ are identical thermalizing channels.}
\label{protocol_fig}
\end{figure}
\emph{Description of the proposed protocol -} Thermometry, in its simplest form, consists of the following. The bath, whose temperature is to be estimated precisely, is at temperature $T$.  A probe is sent to the bath, and then recovered. During this time interval, the probe interacts with the bath, resulting in the final state of the probe imbibing some information about the bath temperature $T$. An estimate of the bath temperature is then obtained by analyzing the probe.  Our protocol is based on the above model with the following crucial difference. The probe interacts with the bath twice in succession, and the ordering between these two interactions is superposed with the help of a control qubit. If the control state is $|0\rangle$, one such ordering is followed, i.e., in Fig.\ref{protocol_fig}, the channel $M$ is encountered first, followed by the identical channel $N$. If the control is at the orthogonal state $|1\rangle$, then the alternative ordering is followed, i.e., the channel $N$ is encountered first, followed by the identical channel $M$. Now, if the control is in a superposition of these two orthogonal states, no specific and definite ordering remains, and the resulting configuration is called a \emph{quantum switch}. For simplicity, we assume that the probe is a qubit, and when in contact with the bath, undergoes thermalization, which can be modelled by a generalized amplitude damping channel $\mathcal{N}$ with the following Kraus operators \citep{nielsen_book}
\begin{widetext}
\begin{equation}
K_{0} = \sqrt{p} \begin{pmatrix}
1  & 0 \\
0 & \sqrt{1 - \lambda}
\end{pmatrix},
K_{1} =  \sqrt{p}\begin{pmatrix}
0  & \sqrt{\lambda} \\
0 & 0
\end{pmatrix}, 
K_{2} =  \sqrt{1-p}\begin{pmatrix}
\sqrt{1 - \lambda}  & 0 \\
0 & 1
\end{pmatrix}, 
K_{3} =   \sqrt{1-p} \begin{pmatrix}
0  & 0 \\
\sqrt{\lambda} & 0
\end{pmatrix}, p = \frac{1}{1 +e^{- \beta \epsilon}} , \lambda = 1 - e^{-t/\tau},
\label{kraus_eq}
\end{equation}
\end{widetext}

\noindent where $t$ is the time of interaction with the bath, $ \epsilon$ is the energy spacing in the probe qubit, $\tau$ is the relaxation time of the bath, and $\beta = 1/T$ is the inverse temperature of the bath. The Kraus operators are normalized, i.e., $\sum_{i} K_{i}^{\dagger} K_{i} = \mathbb{I}$.  If the control state is initially $\rho_c = |\psi_c\rangle \langle \psi_c |$, where $|\psi_c \rangle = \sqrt{\alpha} |0\rangle + \sqrt{1-\alpha} |1\rangle$, then the output state of the correlated  system-control is given by \citep{ebler}
\begin{equation}
\mathcal{E} \ [\rho \otimes \rho_c]= \sum_{i}\sum_{j} W_{ij} (\rho \otimes \rho_{c})W_{ij}^{\dagger},
\label{evolution_eq}
\end{equation}
where $W_{ij} = K_{i} K_{j} \otimes |0\rangle \langle 0| + K_{j} K_{i} \otimes |1\rangle \langle 1|$ are the Kraus operators for the combined probe-control joint system. 

It is clear that tracing out the control from the output state leaves us back with the same state which we would have obtained in the absence of the control.  However, the correlation between the control and the probe established through the thermalizing channel may also store some information about the bath temperature, thus enhancing the precision of estimation of temperature. In the present work, we quantitatively investigate this phenomenon.

\emph{Thermometry with a quantum switch -} For simplicity, we assume that the time spent by the probe inside the bath is much longer than the relaxation time $\tau$ of the bath, or $\lambda = 1 - e^{-t/\tau}$ tends to unity vide \eqref{kraus_eq}. Hence, following \eqref{evolution_eq}, the joint output state of the probe and the control reads as 

\begin{equation}
\rho_{\text{out}} = \begin{pmatrix}
\alpha p & p^2 \sqrt{\alpha(1-\alpha)} & 0 & 0 \\
p^2 \sqrt{\alpha(1-\alpha)}  & \alpha (1-p) & 0 & 0 \\
0 & 0 & p(1-\alpha) & 0 \\
0 & 0 & 0 & (1-p)(1-\alpha) 
\end{pmatrix} . \nonumber 
\end{equation}
\noindent Note that the above density matrix without coherence in the control qubit is a diagonal one, hence the off-diagonal, or genuinely quantum contributions to QFI in \eqref{qfi_defn} does not exist. However, in the case of an initially coherent control qubit, there is a non-zero magnitude of genuinely quantum contribution to the QFI. This affirms that qubit thermometry at equilibrium does benefit from quantum features other than the mere discreteness of levels. The QFI for the output state  above with respect to the parameter $\beta$ is expressed in the following form

\begin{equation}
\mathcal{F}_{\beta} (\rho_{\text{out}}) = \epsilon^2 \frac{ [2+4 \alpha(1-\alpha)]e^{3 \beta \epsilon} + 3 e^{2 \beta \epsilon} + e^{\beta \epsilon} }{(1+ e^{\beta \epsilon})^3 (1+2 e^{\beta \epsilon})}.
\label{qfi_expression_eq}
\end{equation}

\begin{figure}[h]
\includegraphics[scale=0.35]{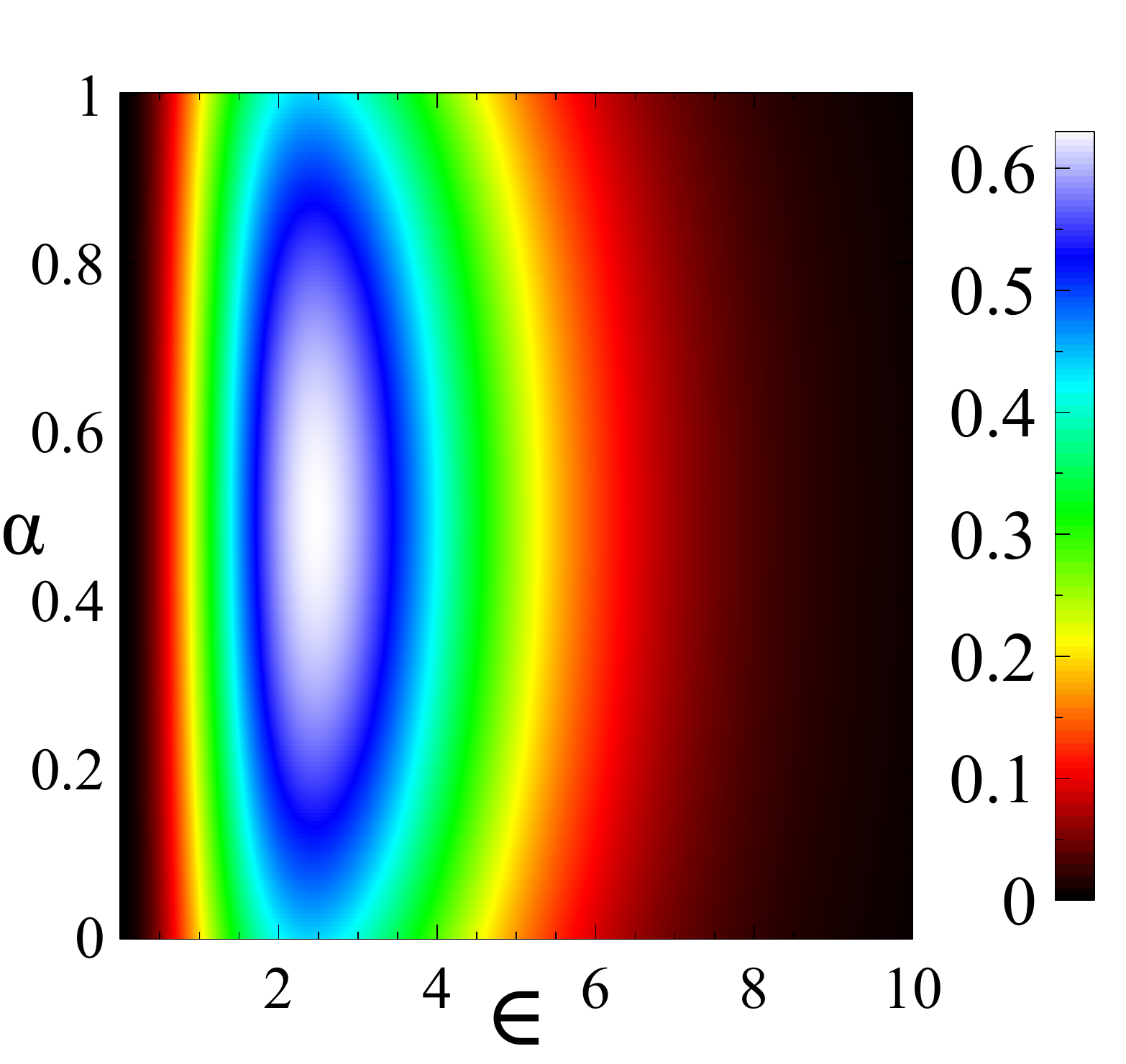}
\caption{(color online) Density plot of QFI for the inverse temperature $\beta$ on the gap $\epsilon$ as well as the superposition parameter $\alpha$. Temperature is fixed at $T=1$.}
\label{gap_3d_fig}
\end{figure}

If we parametrize $\beta \epsilon = \epsilon /T = x$ and optimize over $x$ to maximize the QFI, the corresponding condition is given by $\partial_x \mathcal{F}_{\beta} (x) = 0$, which, upon simplification, yields the following transcendental equation for optimal $x = x^{*}$
\begin{equation}
\xi = \frac{(1+e^{x^{*}})(1+2e^{x^{*}})^2 \left[ (x^{*}-2) e^{x^{*}} - (x^{*}+2)\right]}{e^{2x^{*}} \left[ (2+3x^{*})+(6+4x^{*})e^{x^{*}} + (4-2x^{*}) e^{2 x^{*}} \right]},
\end{equation}
where $\xi = 4\alpha(1-\alpha)$ indicates the amount of superposition initially in the control qubit. In term of resource theory of coherence \citep{baumgratz}, $\xi$ is the square of the $l_1$-norm of coherence $C_{l_1}$,i.e., $\xi = C_{l_1}^{2} (|\psi_c\rangle)$. If $\xi =0$, this indicates the presence of a definite causal order, and $\xi=1$ corresponds to the maximal superposition in the control qubit. For $\xi =0$, the condition above reduces to the optimization condition $\eqref{opt_eqn}$ derived in Ref. \citep{correa_prl}. In case of maximal superposition, i.e., $\xi =1$, the condition above leads to  the solution $x^{*} \approx 2.4741$, which is quite close to the solution of the optimality condition in the presence of a definite causal order. Thus, other things equal, the operating window of the thermometer does not shift much in  presence of the switch. See Fig. \ref{gap_3d_fig} for an illustration of how the QFI depends on the energy gap.

\emph{Thermodynamic uncertainty revisited -} Thermodynamic undertainty relations have a long history, which we shall not dwell upon here \citep{tur1,tur3}. We note that the thermodynamic uncertainty relation in \eqref{tur} is analogous to the familiar uncertainty relation for incompatible observables. It is well known \citep{berta,arunwilde} that the presence of quantum entanglement and quantum superposition can reduce uncertainty. Thus, it is natural to wonder whether the quantum switch induces a similar effect for the thermodynamic uncertainty relation. Indeed, starting from \eqref{qfi_expression_eq}, and the quantum Cramer Rao bound, yields the following version of the thermodynamic uncertainty relation
\begin{equation}
\delta \beta \Delta H \geq   \frac{1}{\sqrt{1 + \frac{\xi \epsilon^2}{(1+e^{-\beta \epsilon})(2+e^{-\beta \epsilon})}}}.
\end{equation}

\noindent Since the quantum Cramer Rao bound for estimating a single parameter is tight, it is possible to saturate the above uncertainty relation as well.  We now concentrate on limiting  cases. If the bath temperature is very low, i.e., $\beta = 1/T \rightarrow \infty$, this reads as 
\begin{equation}
\delta \beta \Delta H \geq   \frac{1}{\sqrt{1 + \frac{\xi \epsilon^2}{2}}}.
\end{equation}
\noindent On the other hand, if the bath temperature is very high, i.e., $\beta = 1/T \rightarrow 0$, the corresponding thermodynamic uncertainty relation is given by
\begin{equation}
\delta \beta \Delta H \geq   \frac{1}{\sqrt{1 + \frac{\xi \epsilon^2}{6}}}.
\end{equation}
\noindent  The lower bound depends on the quantum coherence of the control qubit and the energy gap of the probe qubit. Therefore, one can see that similar to earlier results \citep{berta,arunwilde}, quantum coherence reduces thermodynamic uncertainty. Also, it is evident from the above that a large gap in the probe Hamiltonian reduces the thermodynamic uncertainty. However, it is easy to see that the average energy of the probe in such a case becomes very big. Thus, the assumption that the probe is much smaller in comparison to the bath may no longer hold.  
\begin{figure}[h]
\includegraphics[scale=0.25]{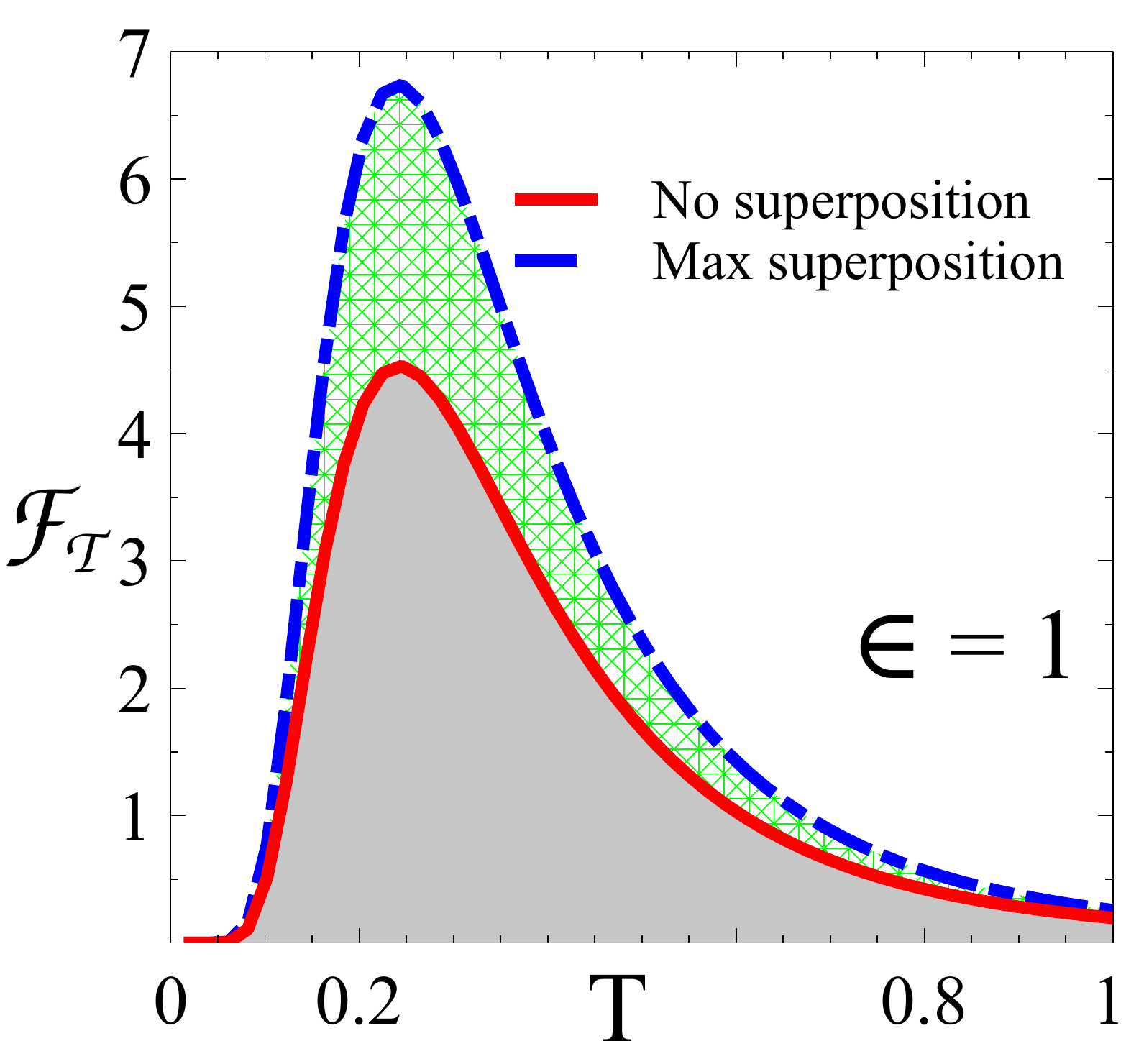}
\includegraphics[scale=0.25]{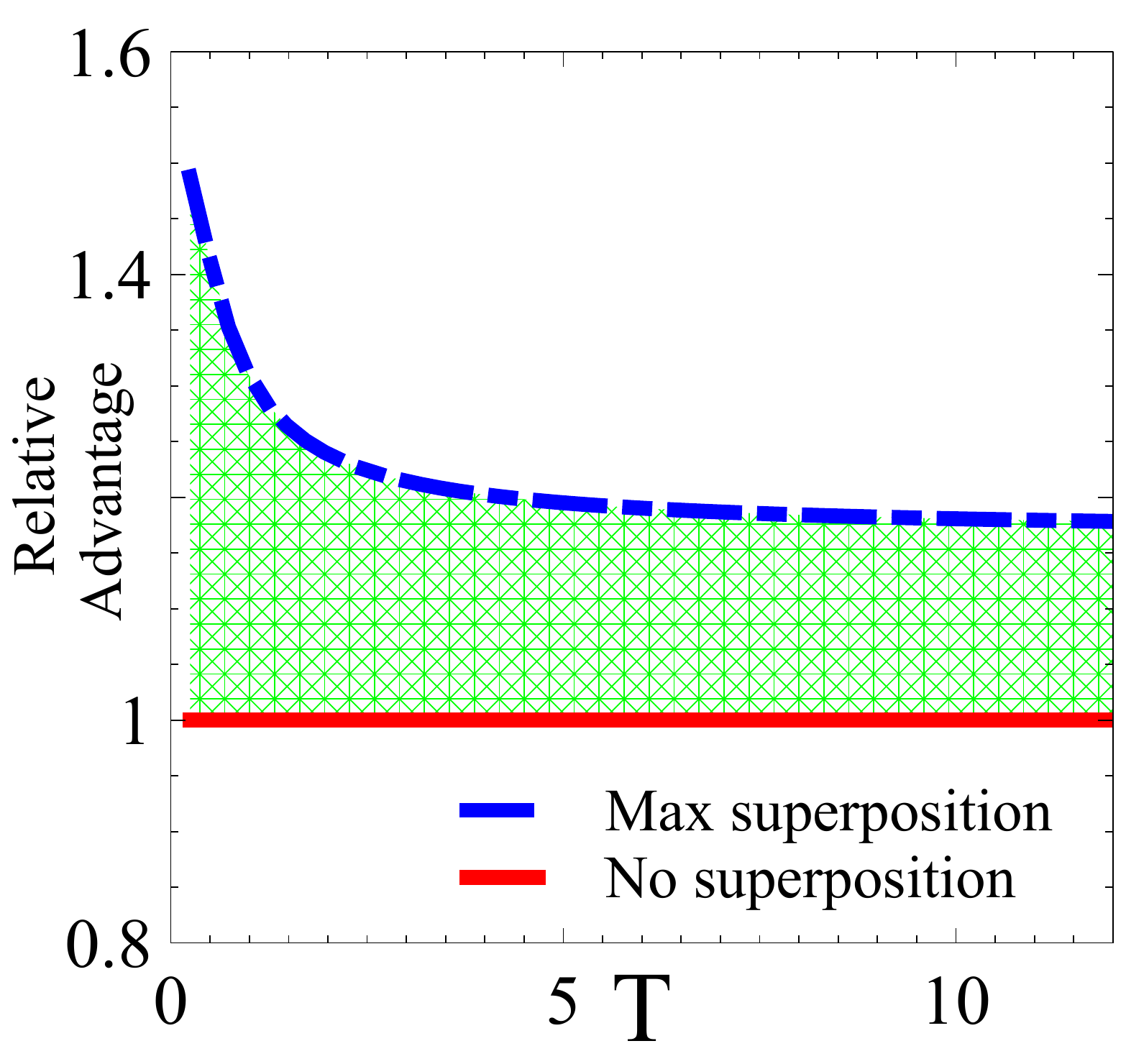}
\caption{(color online) \emph{Left}: QFI for temperature $T$ vs. temperature $T$. \emph{Right}: Relative advantage in terms of QFI gained through the use of quantum switch. The red solid line indicates the qubit probe used without any switch, the blue dash-dotted line indicates the qubit probe used with a maximally coherent switch. Green shaded regions are only accessible when using a quantum switch in the qubit protocol.}
\label{2d_fig}
\end{figure}

\emph{Performance advantage -} Let us now quantify the relative advantage gained through the use of the qubit probe with a quantum switch vis-a-vis a conventional qubit probe. The relative gain in QFI through the use of a quantum switch utilizing a  maximally coherent control qubit, with respect to a conventional qubit probe, reads as 
\begin{equation}
\chi = \frac{\mathcal{F}_{\beta}^{\text{switch}}}{\mathcal{F}_{\beta}^{\text{no switch}}} = \frac{ (2+\xi)e^{3 \beta \epsilon} + 3 e^{2 \beta \epsilon} + e^{\beta \epsilon}}{ 2 e^{3 \beta \epsilon} + 3 e^{2 \beta \epsilon} + e^{\beta \epsilon}}.
\end{equation}
\\ 
In the limit of very high temperature, i.e., $\beta \rightarrow 0$, and maximal superposition between paths, the ratio $\chi \approx 7/6$, whereas in the limit of very low temperature, i.e., $\beta \rightarrow \infty $ and maximal superposition between paths, the ratio $\chi \approx 1.5$. Expressed in terms of precision of estimation of temperature, this translates to $\approx 8\%$ more precision for estimating a very high temperature, and $\approx 22\%$  more precision for estimating a very low temperature. Thus, our protocol performs much better than the other qubit thermometry protocols in the low-temperature regime, while retaining the advantage vis-a-vis other protocols in the high-temperature regime as well. See Fig. \ref{2d_fig} for an illustration of the advantage gained through the use of the quantum switch.

\emph{Comparison with a Harmonic Oscillator probe - } A qubit has only two energy levels, therefore the problem of optimizing the Hamiltonian spectrum does not arise in general except optimizing over the value of the energy gap. Extending the optimal thermometry scheme \citep{correa_prl} for $N$-level systems leads us to an optimal Hamiltonian spectrum with a gapped ground state and a $N-1$ fold energetically degenerate energy eigenstates. Clearly, designing such probes is practically quite challenging. In contrast, Harmonic oscillators with equispaced energy levels are far more accessible. It was shown \citep{correa_prl} that they are superior to qubit probes with the same energy gap $\epsilon$ in terms of precision as well as the breadth of the operating window. The corresponding QFI for the conventional Harmonic oscillator probe is given by \citep{sanpera_review}

\begin{equation}
\mathcal{F}_{\beta}^{\text{HO}} = \epsilon^2 \frac{e^{-\beta \epsilon}}{(1-e^{-\beta \epsilon})^2}
\label{ho_eqn}
\end{equation}

\begin{figure}[h]
\includegraphics[scale=0.35]{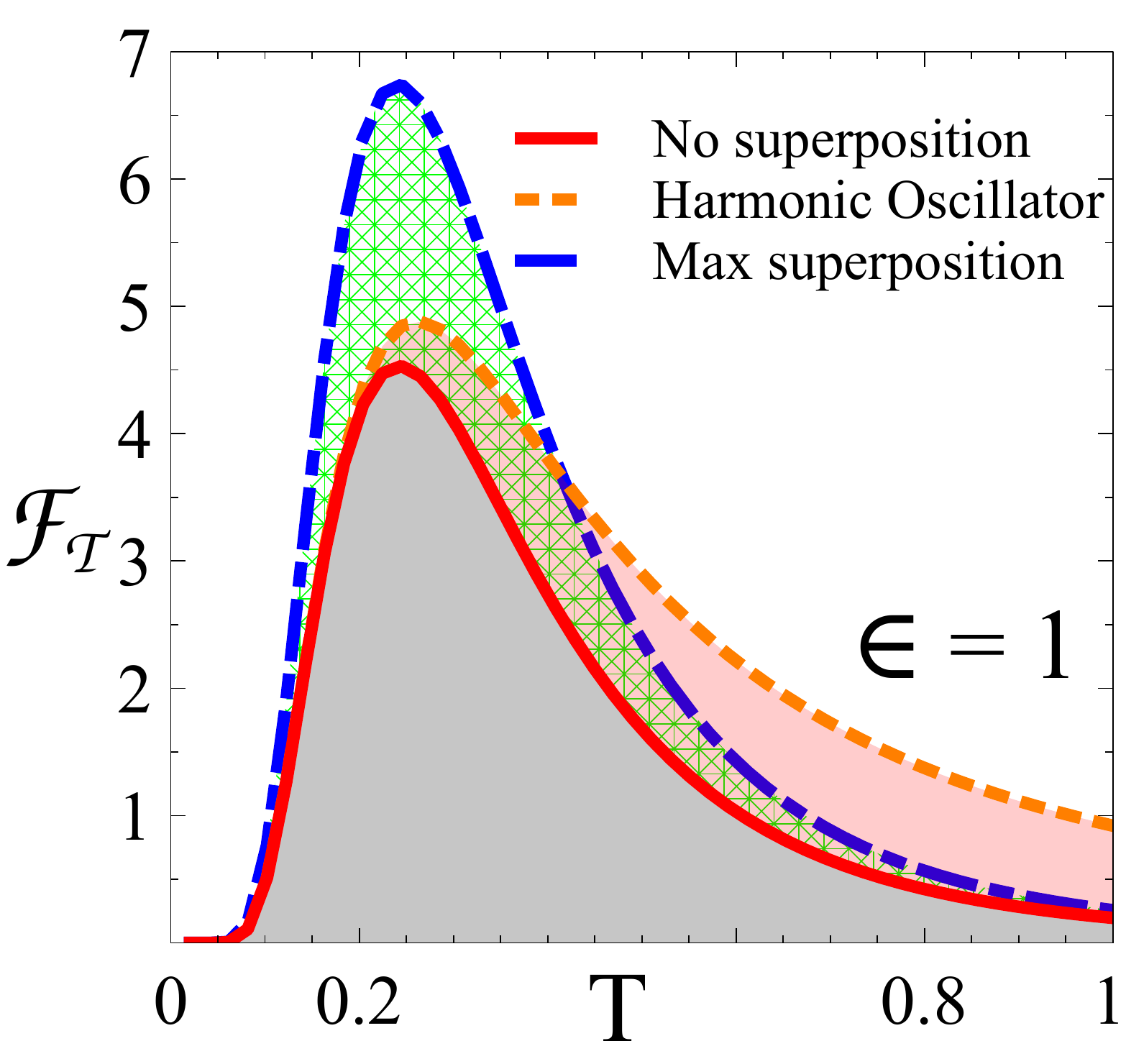}
\caption{(color online) Comparison of the QFI for temperature with temperature $T$ for the qubit probe without quantum switch (red solid line), qubit probe with a maximally coherent quantum switch (blue dash dotted line), and a harmonic oscillator probe (orange dotted line). The energy level spacing in every case is $\epsilon = 1$.}
\label{ho_fig}
\end{figure}

Comparing \eqref{ho_eqn} with the expression \eqref{qfi_expression_eq} of QFI for a qubit probe with a quantum switch reveals an interesting phenomenon. Below a certain threshold temperature, our qubit probe  with a maximally coherent control outperforms the conventional Harmonic oscillator probe. Even better, this region includes the operating ranges of our probe as well the conventional harmonic oscillator probe. See Fig \ref{ho_fig} for illustration. Thus, it is actually better to use a qubit probe in conjunction with a quantum switch rather than a conventional Harmonic oscillator probe, even though the latter has an infinite number of energy levels. The threshold temperature is obtained from equating the QFI expressions for the qubit probe with a quantum switch \eqref{qfi_expression_eq}, and the conventional harmonic oscillator probe \eqref{ho_eqn}, which yields the following equation for $\beta \epsilon = x$
\begin{equation}
\frac{ 3 e^{3 x} + 3 e^{2 x} + e^{x}}{(1+ e^{x})^3 (1+2 e^{x})} = \frac{e^{-x}}{(1-e^{-x})^2}.
\end{equation}
This is an algebraic quartic equation in $e^{x}$ and can be shown to have a non-zero real solution of $x \approx 2.40 $. For example, assuming $\epsilon =1$ yields the threshold temperature to be $T_{\text{threshold}} \approx 0.4157 $, which is not in the optimal temperature window for our scheme. 

\emph{Conclusions and Outlook-} We have proposed a protocol for qubit thermometry using a quantum switch and shown that this protocol allows for significantly more precise thermometry than the \emph{optimal} qubit probe considered so far, and even surpasses the precision offered by a Harmonic oscillator probe. We have confined ourselves to simplest, i.e., qubit probes in the present work. A direct generalization to $N$-level or continuous variable probes remains to be explored. In this work, we have considered only the equilibrium configurations. It remains to be seen whether the quantum switch offers additional advantages in the transient regime, which is practically relevant for baths with a very long relaxation time compared to the time allowed for thermometry. Finite dimensional probes usually offer a narrower operating window than Harmonic oscillator type probes. It is important to consider whether using a quantum switch can increase the operating window as well. Lastly, if one considers a quantum switch with many, and not just two possible pathways, then it is of vital interest to know how the quantum advantage scales with the number of possible pathways.  More generally, our results open up the possibility that the use of quantum switches would give rise to more practical applications by optimizing several metrological tasks.  From a resource theoretic perspective \citep{manab}, it is ultimately the coherence in the control state and the subsequent superposition of configurations which allows for the observed advantage. We hope that quantum switch will be considered as another tool in addition to quantum coherence and entanglement in quantum metrology.

\emph{Acknowledgment-} CM acknowledges a doctoral research fellowship from the Department of Atomic Energy, Government of India, as well as funding from INFOSYS. MKG acknowledges a post-doctoral fellowship from the Department of Atomic Energy, Government of India.
\bibliographystyle{apsrev4-1}
\bibliography{thermometry_project}

\end{document}